# Increase of cationic concentration due to bending of overcharged DNA in strong Coulomb coupling regime.

Arup K mukherjee

Department of Physics, Chancellor College, Box 280. Zomba, University of Malawi, Malawi

E-mail: akmukherjee11@hotmail.com

## Abstract

This study reveals that, in strong coulomb coupling regime, bending a straight and fully overcharged DNA (up to its 'maximal acceptance' by multivalent cations) to a circle releases some of the adsorbed (correlated) cations but still remains fully overcharged.  This phenomenon seems to be inherent to the minimum energy state of a DNA. By definition, the total electrostatic potential energy of a macroion-counterion system reaches to its lowest point at maximal acceptance of overcharging counterions that ensures the most stable conformation. This intermediate phenomenon of release of cations from DNA surface due to bending can be taken into account in theoretical modeling of some ionic concentration dependent physico-chemical aspects of DNA solutions in strong Coulomb coupling regimes.

## 1. Introduction

The counterintuitive phenomenon of charge reversal or overcharging of charged macroions in solutions containing counterions is an experimental fact [1-3]. Theoretical and computational approaches explaining the phenomenon are abundant in literature [4-23]. It is understood that under specific solution conditions overcharging is a natural phenomenon. The dominant driving mechanism for overcharging is the special correlation which builds up among the counterions attached with the oppositely charged macroion surface. These counterions are then no longer free to move in solution. Due to this correlation counterions can accumulate on the macroion surface in such an amount whose total charge exceeds the bare charge of the macroion. The dielectric constant of the solution is one of the important factors that control the efficiency of overcharging. The total electrostatic potential energy among charges depends basically on the dielectric constant. Solution with low dielectric constant is generally termed as strong Coulomb coupling where the total electrostatic potential energy can be much higher than thermal energy. In other words, the strength of Coulomb coupling depends on the value of Bjerrum length $l_B$ defined as $l_B = e^2/4\pi\varepsilon_o\varepsilon_r k_B T \Rightarrow e^2/\varepsilon k_B T$, where $\varepsilon$ is the dielectric constant and T is absolute temperature. Even though the dielectric constant is a characteristic of a pure solvent it can be decreased by mixing with other compounds such as ethanol [27]. For the present study a fixed temperature of 275 K and a dielectric constant of 20 of the solvent (water solution with ethanol and counterions with no added salt) has been considered ($l_B$ ~ 3.04 nm) [24]. This solution condition can also maintain the strong Coulomb coupling environment in liquid water.

This study reveals that Overcharging causes a DNA to bend rapidly and if a maximally overcharged DNA bends to a circle it releases some strongly attached (correlated) counterions from its surface while maintaing its maximum overcharged state. Thus if one considers a cationic solution (without any buffer salt, for simplicity, and with low dielectric constant) containing straight DNAs then those DNAs are supposed to be overcharged as DNAs are always naturally overcharged in strong Coulomb coupling regime. Due to overcharging the number of multivalent cations (free in solution) must decrease at first. Next when the DNAs bent (to a circle) some of those cations get beck to the solution from the DNA surface and again increase the concentration of the free multivalent cations in the solution. This so far unnoticed phenomenon, a fluctuation in cationic consentration at an intermediate stage, may be considered in theoretical modeling as a transient picture of many biophysical and biochemical interactions. For example, overcharging plays a key role in one of the mechanisms of like-charge attraction [5, 6, 17] which is thought to be responsible for bundle formation/aggregation of like-charged biomolecules (such as DNAs) in cationic solution. It is found that when cationic concentration is further increased beyond a threshold value the reentrance of DNAs in solution occurs [25]. In both aggregation and reentrance events [25-27] the release of cations due to bending of overcharged DNAs may play a role as it causes a fluctuation in cationic concentrations especially in cases of appreciable DNA concentration in strong Coulomb coupling environments.

## 2. Model and Simulation Methods

The system considered in this study is comprised of a cylinder (macroion) of length ($L$) 51 nm and of radius (r) 1 nm (mimicking a c-DNA) with bare macroion charge $Q = -Z_m |e|$ surrounded by a number $N_C$ of small spherical counterions with charge $q = Z_C |e|$ and radius ($R_C$) 0.18 nm so that $Z_m = Z_C No = 300$ in the neutral state. $Z_m$ and $Z_C$ are the macroion and counterion valances respectively. The charge of the cylinder has been considered as being comprised of very closely distributed point charges of magnitude $z_i |e|$ in a line along the axis of the cylinder so that $Z_m = \sum_{i=1}^{n+1} z_i$, where n is the total number of such points. This has been considered instead of a continuous line charge (Manning conception) [30] to facilitate the calculations by avoiding frequent solutions of generally non-complete elliptic integrals [24]. To calculate the total minimized electrostatic energy a previously developed [5] energy minimization simulation technique has been employed. The energy minimization technique is a simulation technique that calculates the counterion positions on the surface of the macroion by minimizing the distances among them for which the total electrostatic potential energy of the macroion-counterion system yields very near to the lowest possible (ground state) energy. For overcharging additional counterions (in excess to the neutral state) are added one by one and after each addition the corresponding minimum total energies are calculated by minimizing their positions. Here one needs to consider all the counterions are always at a constant counterion-macroion distance ($\tau = r + R_C$) of closest approach. This is an intrinsic requisite of the technique. This condition is also required to maintain the environment of strong Coulomb coupling [20,21]. As the macroion has been considered as hard core the Lennard-Jones potential calculations are not required for this study. Note that it has been shown earlier [5] that the energy minimization technique produces exactly the same results as MD or MC under strong Coulomb coupling condition. But this technique is rather simple and easy to use (without considering a cell or a solution surrounding the DNA) for any macroion geometry. The technique converses rapidly and thus economic in terms of computer time.

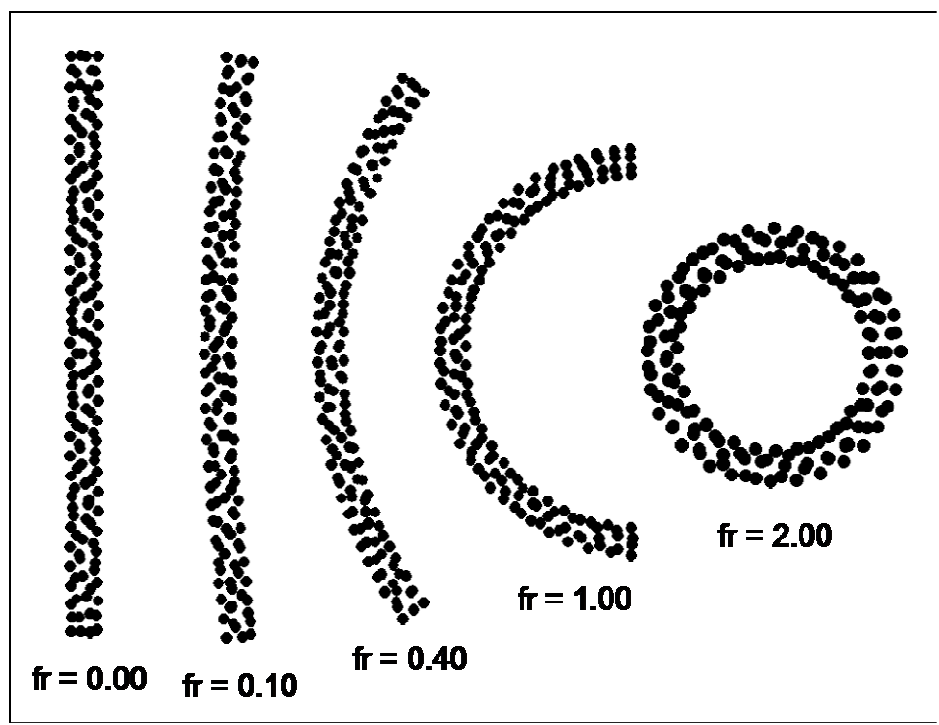

**Figure 1.** The process of gradual bending of a straight cylinder (surrounded by energy minimized counterions of any valance) to a circle. The black dots represent counterions. 'fr' is the bending fraction that varies from 0 (straight line ) to 2 (circle).

For the purpose of the present study, a straight cylinder has been considered at first and the total minimized energy (counterion-counterion plus counterion-macroion for a certain $Z_c$) has been calculated at its neutral state. Then the overcharging has been performed for $N_{max}+1$ number of additional counterions, where $N_{max}$ is termed as 'maximal acceptance' which is the number of counterions that yields the lowest energy (see figure 5). Next, the cylinder has been bent a little to an arc of a circle and the minimized energy has been measured by the same way as in the case of the straight cylinder. The process of bending continues until the cylinder forms a circle. In each step of bending (fr) a curve (like figure 5) is achieved which consists of a set of points (degrees of overcharging) ranging from zero (neutral) to $N_{max}+1$. These are termed as overcharging curves.

The bending fraction 'fr' is defined as $L/\pi R$, where R is the radius of curvature. Obviously 'fr' can take up any value between zero and two. Figure 1 depicts the process of bending.

The total electrostatic energy of the counterion-macroion system reads,

$$E_{Coul} = \frac{|e|^2}{4\pi\varepsilon_o \varepsilon_r} \left[ \sum_{i=1}^{n} \sum_{j=1}^{m} \frac{z_j Z_C}{\tau_{ij}} + \sum_{i<j} \frac{Z_C^2}{r_{ij}} \right] \qquad (1)$$

where $\varepsilon_r$ is the relative permittivity, $\tau_{ij}$ is the distance between any counterion i and a point macroion charge j and $r_{ij}$ is the separation between any two counterions i and j. n varies as $N_o \leq n \leq N_{max} + 1$. Where $N_o$ is the neutral state counterion number. Equation 1 can be written as

$$E_{Coul}/k_B T = l_B \left[ \sum_{i=1}^{n} \sum_{j=1}^{m} \frac{z_j Z_C}{\tau_{ij}} + \sum_{i<j} \frac{Z_C^2}{r_{ij}} \right] \qquad (2)$$

It is worth to mention that the relative permittivity of the cylinder mimicking the c-DNA has been considered identical to that of its out side (the surrounding counterions) to avoid image charge problems. The Bjerrum lelgth has been taken as 30.38 Å all over the study. As the simulation technique converges significantly rapidly, around 300,000 moves per counterion is sufficient to reach very close to the lowest possible energy state.

## 3. Modified Scatchard Model

A simple theoretical model to fit the simulation data for all geometries has been proposed by modifying [4] the Scatchard [28] approach, which employs average interactions. For the macroion complexes (with N counterions) of any geometry, the average interaction can be expressed as

$$\langle E_N \rangle = Z_C^2 \langle C \rangle \left[\frac{N(N-1)}{2}\right] - Z_m Z_C \langle M \rangle N$$

$$\langle E_N \rangle = Z_C^2 \langle C \rangle \left[\frac{(N_o+n)(N_o+n-1)}{2}\right] - Z_m Z_C \langle M \rangle (N_o + n) \quad (3)$$

where $N = N_o + n$, $N_o = Z_m/Z_C$ and n is the overcharging counterions. $\langle C \rangle$ and $\langle M \rangle$ represent average counterion-counterion and counterion-macroion energy functions. The above equation can be expressed in quadratic form in terms of n as

$$\langle E_N \rangle = S_o + S_1 n + S_2 n^2 \quad (4)$$

where

$$S_o = Z_C^2 \langle C \rangle \left[\frac{N_o(N_o-1)}{2}\right] - Z_m Z_C \langle M \rangle N_o$$

$$S_1 = \frac{Z_C^2}{2}[2N_o(\langle C \rangle - \langle M \rangle) - \langle C \rangle]$$

$$S_2 = \frac{Z_C^2}{2}\langle C \rangle$$

The energy difference between a neutral complex and an overcharged on is

$$\Delta E_n = \langle E_N \rangle - \langle E_{N_o} \rangle$$

$$= S_1 n + S_2 n^2 \qquad (5)$$

Using the first overcharge $\Delta E_1$ from the simulation data, $\langle C \rangle$ can be calculated from the above equation as

$$\langle C \rangle = \frac{\Delta E_1}{Z_C^2 \left[1 - \frac{\langle M \rangle}{\langle C \rangle}\right] N_o} \qquad (6)$$

Thus from (5) and (6) one can write

$$\Delta E_n = \frac{\Delta E_1 n}{2\left[\frac{\langle M \rangle}{\langle C \rangle} - 1\right] N_o} \left\{ 2\left(\frac{\langle M \rangle}{\langle C \rangle} - 1\right) N_o + 1 - n \right\}$$

$$= \frac{\Delta E_1 n}{X N_o} \{X N_o + 1 - n\} \qquad (7)$$

where $X = 2\left\{\frac{\langle M \rangle}{\langle C \rangle} - 1\right\}$ is assumed as an arbitrary fit parameter.

The 'maximal acceptance' $n_{max}$ can be calculated by maximizing $\Delta E_n$ (equation (5)) as

$$\frac{\partial (\Delta E_n)}{\partial n} = S_1 + n\left(\frac{\partial S_1}{\partial n}\right) + 2 S_2 n + n^2 \left(\frac{\partial S_2}{\partial n}\right) \qquad (8)$$

But $\qquad \frac{\partial S_1}{\partial n} = \left(\frac{\partial S_1}{\partial \langle C \rangle}\right)\left(\frac{\partial \langle C \rangle}{\partial N_{max}}\right)\left(\frac{\partial N}{\partial n}\right)$

Since $N = N_o + n$

$$\frac{\partial S_1}{\partial n} = \left(\frac{\partial S_1}{\partial \langle C \rangle}\right)\left(\frac{\partial \langle C \rangle}{\partial N}\right)$$

Using these, equation (8) can be written as

$$\frac{\partial (\Delta E_n)}{\partial n} = S_1 + 2 S_2 n + (2 N_o - 1)\frac{Z_n^2}{2}\left(\frac{\partial \langle C \rangle}{\partial N}\right)$$

For a large number of counterions, one can assume that $\frac{\partial \langle C \rangle}{\partial n} = 0$, since the inclusion of one ion does not make a significant change in the total energy if the counterion number is large enough. This approximation is within the frame work of the Scatchard approach. Then the localization of the minimum in the $\Delta E_n$ profile is

$$n_{max} = -\frac{S_1}{2S_2}$$

$$= N_o\left[\frac{\langle M \rangle}{\langle C \rangle} - 1\right] + \frac{1}{2} \qquad (9)$$

$$= \frac{XN_o + 1}{2} \qquad (10)$$

Thus from equation (10) $n_{max}$ can be calculated from the fit parameter X. The nearest integer of $n_{max}$ is the 'maximal acceptance'.

## 4. Results and discussion

The overcharging curves (OC) for multivalent counterions are shown in figure 2. The solid and broken lines are from equation (7). The fit parameters and the first overcharge $\Delta E_1$ of each curve are given in table 1. Figure 2 also shows a comparison between OCs of circular and straight (rod) DNAs in a familiar way [4,5] where neutral state energies (which are different for different valence counterions) have been subtracted from the energies of all degrees of overcharges. Due to this, the OCs of the circular DNAs for all valances lie above the OCs of straight DNAs since the neutral state energies of circular DNAs are always lower than those of straight DNAs. (Practically this picture looks like opposite to Figure 3) . Figure 3 shows the exact positions of those curves which have been plotted without any subtractions. It depicts clearly that the energies of circular DNAs are much lower than those of straight DNAs for all valance of overcharging counterions and for all degrees of overcharging. It also shows that overcharging is energetically more favorable for bent morphology than its straight shapes.

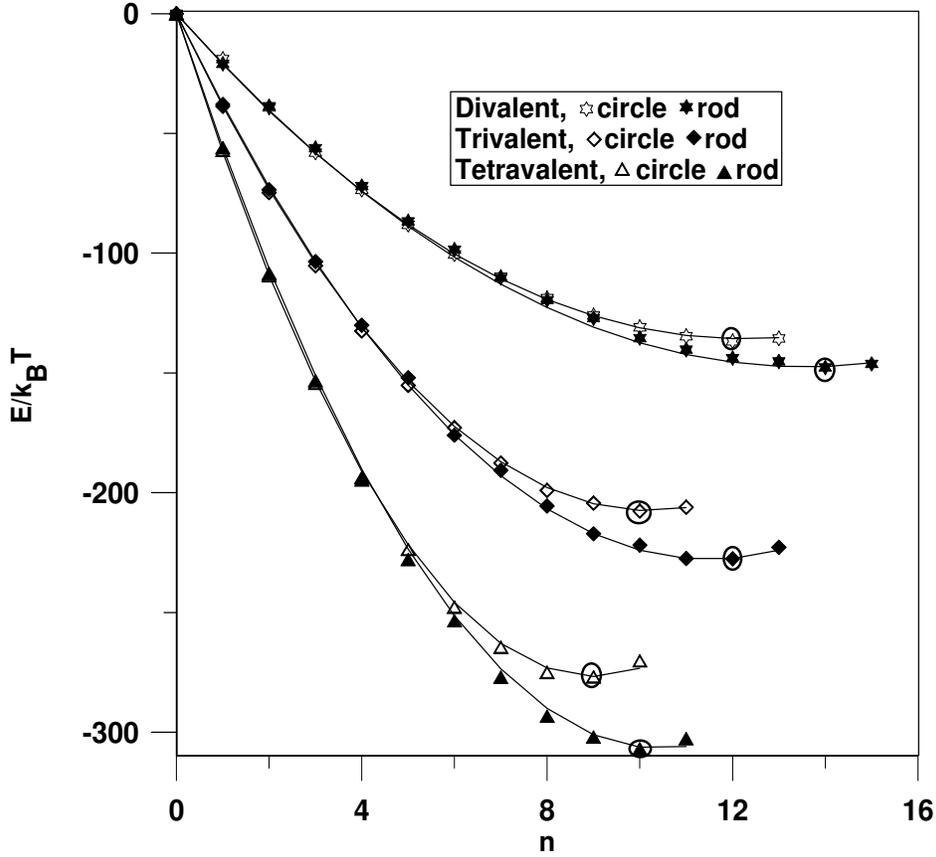

**Figure 2.** A conventional way of comparison between a straight (rod) and a circular DNA (510 Å) overcharging energies. For each curve the corresponding neutral state energy has been taken as the reference point. n is the number of overcharging counterions. Thus neutral state corresponds to n = 0. The $n_{max}$ of all curves are indicated by circles. The solid lines are from the equation (7). The fit parameters are given in table 1.

In figure 2, $N_{max} = N_o + n_{max}$, $N_o = Z_m/Z_c$. $N_{max}$ represents the number of counterions for which the total electrostatic energy is the lowest for a specific valance. $N_{max}$ of each curve are indicated by a circle in figure 2. The most interesting feature of this study is that the $N_{max}$ of circular DNAs are always less than that of straight (rod) DNAs. This implies that when a maximally overcharged DNA bends to a circle it rejects some counterions but still remains maximally overcharged. The number of rejected counterions varies with valance. For example, for the di- and trivalent counterions the rejected number is two, while for tetravalent it is one (see figure 2). Due to this rejection the cationic concentration of the solution can increase which can, in turn, change dramatically the whole chemical picture of the DNA solution.

The overcharging curves for other bending fractions $0 < fr < 2$ have been seen to very similar to those shown in figure 2 and been identified in between fr = 0 and fr =2 curves and thus have not shown.

From the modified Scatchard model using the values of the fit parameters X one can calculate the 'maximal acceptance' employing equation (10). The results are tabulated in table 1. The results are in complete agreement (see fig. 2) with the simulation data given in table 2.

**Table 1**. Modified Scatchard Model (equation (6)) fit parameters for figure 2.

| Valance | Straight | | | Circular | | |
|---|---|---|---|---|---|---|
| | $\Delta E_1$ | X | $N_{maxc}$ | $\Delta E_1$ | X | $N_{maxc}$ |
| 2 | -20.96 | 0.1741 | 164 | -21.22 | 0.1570 | 162 |
| 3 | -37.92 | 0.2220 | 112 | -39.62 | 0.1940 | 110 |
| 4 | -55.98 | 0.2630 | 85 | -57.99 | 0.2270 | 84 |

(Note: $N_{max} = N_o + n_{max}$ have been calculated from equation (10))

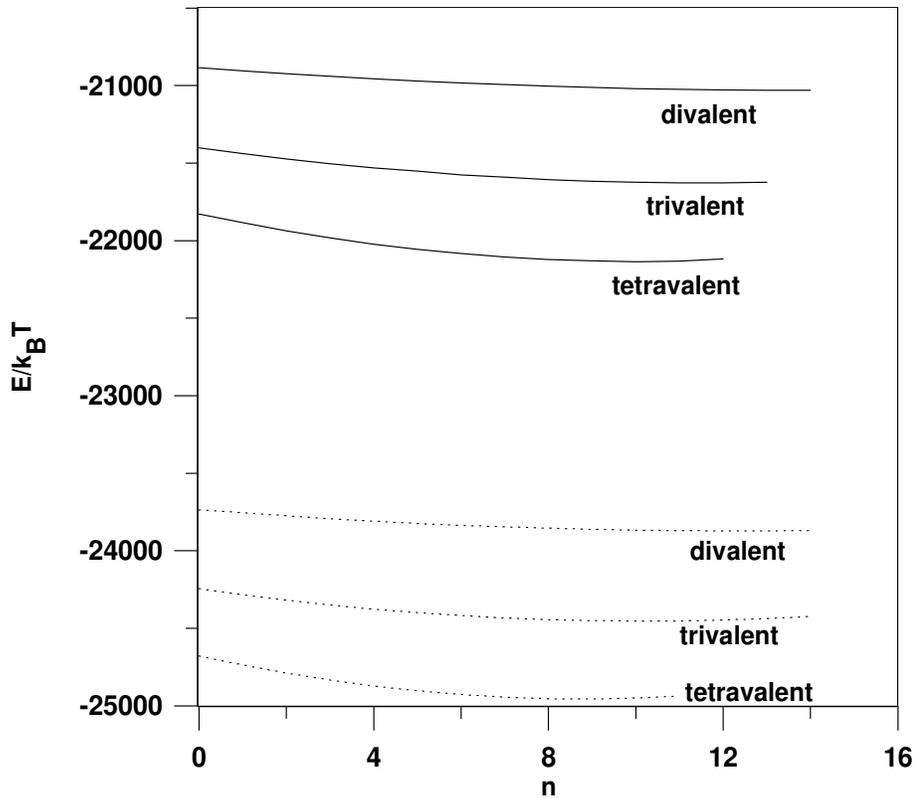

**Figure 3.** A simple and useful way of comparison between a straight (rod) and a circular DNA overcharging energies where no reference point has been considered. n is the number of overcharging counterions. The solid and broken lines (polynomial fits) represent the straight and circular DNAs respectively.

The minimized energies at $N_{max}$ of OCs for various bendings are seen to decrease with bending as shown in figure 4. Filled symbols on the solid curves represent the minimum possible energies corresponding to $N_{max}$ for every degrees of bending and valance. For comparison Purpose the neutral state energies (no overcharging [24]) have also been plotted (open symbols and broken lines). Reasonably the solid curves always lie below the corresponding broken curves which indicate that overcharged states are energetically more stable than their neutral states. Thus a DNA (with any degree of bending) in solution with multivalent counterions becomes naturally overcharged for stability and also assumes circular or other bent morphology than to remain straight. The figure shows that the larger the valance of the counterion the bigger the possibility of bending and overcharging.

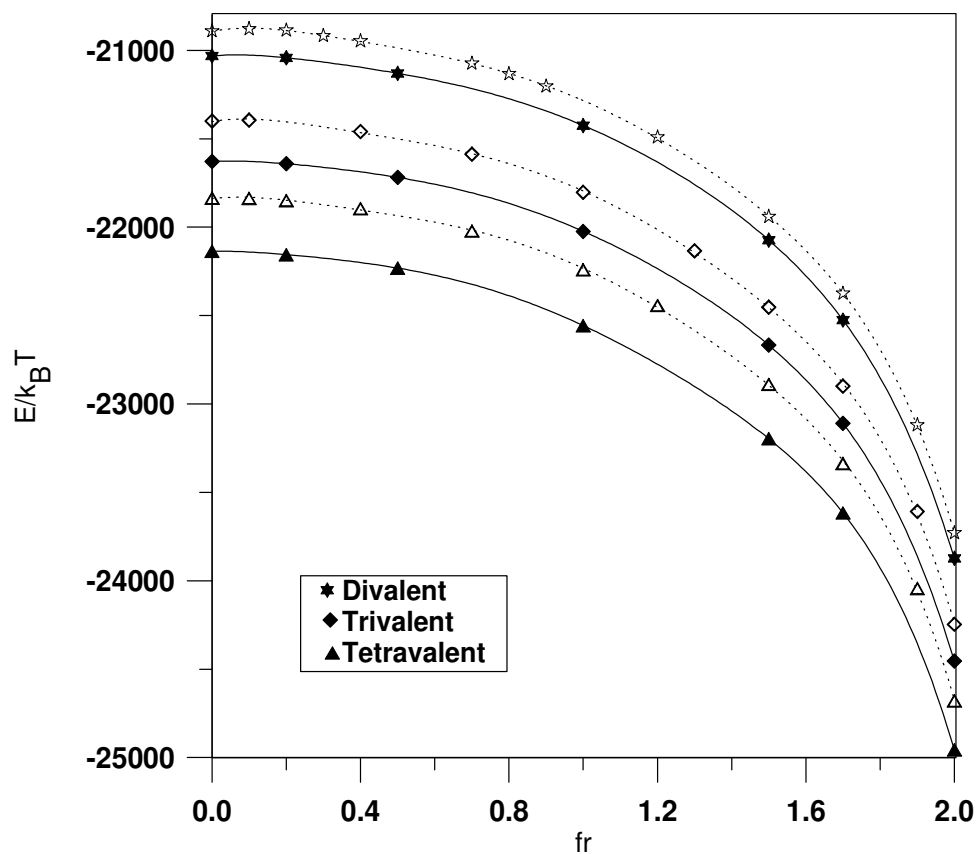

**Figure 4.** Decrease in total electrostatic potential energy with bending. Each filled symbol represents the energy at maximum possible overcharge. The corresponding open symbols represent energies of neutral states [24]. The solid and broken lines are polynomial fits to guide the eyes.

The energy minimized counterion positions on the overcharged straight cylindrical macroion (DNA) surfaces have been shown in figure 5, where usual helical patterns of distributions are observed for different multivalent counterions. In figure 5(a) there are 164 (= $N_{max}$) divalent Counterions, 14 (=

$n_{max}$) excess to its neutral state. In figure 5(b) and 5(c) the total number of excess tri- and tetravalent counterions are 12 and 10 respectively.

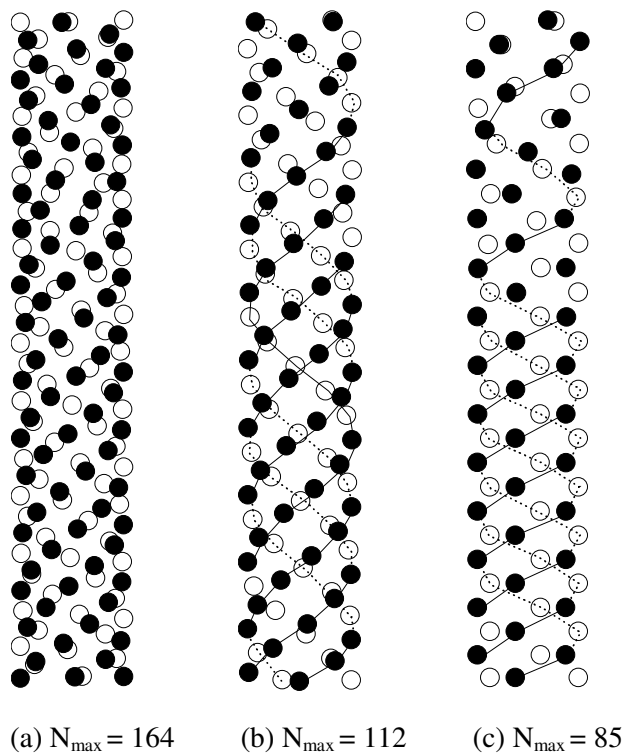

(a) $N_{max}$ = 164   (b) $N_{max}$ = 112   (c) $N_{max}$ = 85

**Figure 5.** Overcharged straight DNAs (fr = 0) of length L = 510 Å. The total number of counterions (maximal acceptances $N_{max} = N_o + n_{max}$) of different valences are given under each figure. The overcharging counterions are (a) divalent (b) trivalent and (c) tetravalent. The black dots represent the counterions those are in the front and the circles represent those are in the back side of the DNAs. The solid and broken lines are drawn to show some of the counterion distribution patterns.

The distribution patterns of multivalent counterions on curved overcharged macroions ($fr = 0.5$) at their maximal acceptances are shown in figure 6, where, the helical patterns are still obvious. It has been observed that for multivalent counterion distributions, there is hardly any tangible change in the usual helical distribution of the counterions due to either any degrees of the bending from a straight cylinder up to its circular form or overcharging up to its maximal acceptance $N_{max}$. After energy minimization always the counterions arrange themselves in helical patterns.

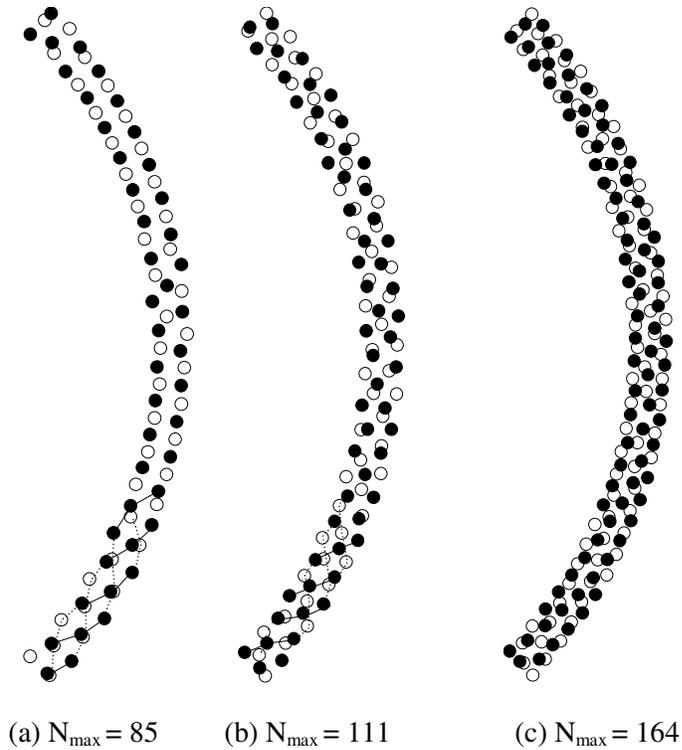

(a) $N_{max} = 85$  (b) $N_{max} = 111$  (c) $N_{max} = 164$

**Figure 6.** The multivalent counterion distributions over the DNA with fr = 0.5. The total numberof counterions (overcharged) are given under each figure. . The overcharging counterions are (a) tetravalent (b) trivalent and (c) divalent. $N_{max}$ is the fully overcharged state ( maximal acceptance) counterions. The black dots represent the counterions those are in the front and the circles represent those are in the back side of the DNAs. The solid and broken lines are drawn to show some of the counterion distribution patterns.

The multivalent counterion distribution patterns on the complete circular form of the overcharged DNAs are shown in figure 7 for all valences. As stated earlier, one can see that the maximal acceptances here are somewhat different from those of figure 5 for their respective valances. These are summarized in table 2. The helical patterns are still present.

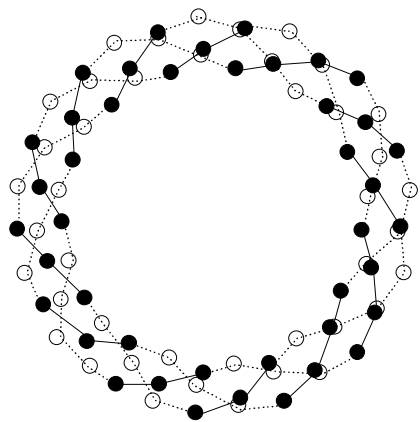

(c) $N_{max} = 84$

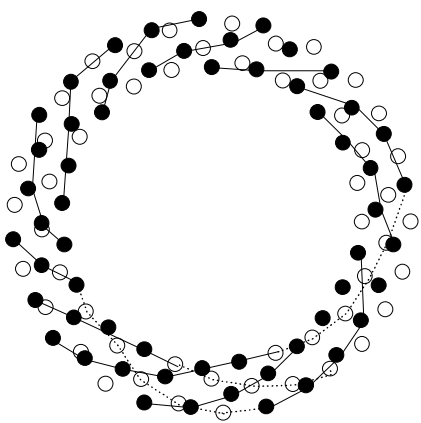

(b) $N_{max} = 110$

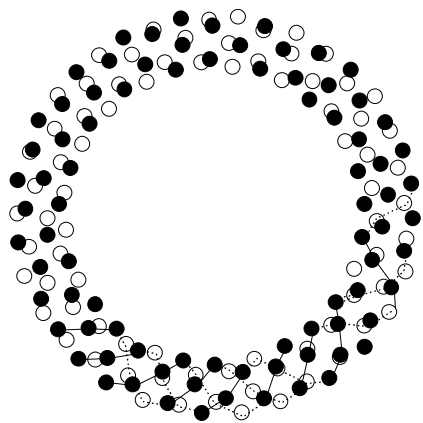

(a) $N_{max} = 162$

**Figure 7.** The distribution of overcharging counterions at $N_{max}$ (solid and open circles) on a circular form of DNA minimizing the total electrostatic energy. $N_{max}$ is the number of counterions of fully overcharged state ( maximal acceptance). The overcharging counterions are (a) divalent (b) trivalent and (c) tetravalent. Solid circles are on the front and the open circles are on the back sides of the macroions. Solid and broken lines were drawn to show the distribution patterns.

**Table 2**. The total electrostatic potential energies, $E_{N_{max}}$, of the straight and the circular DNAs and the corresponding 'maximal acceptances' $N_{max}$ for different valances.

| | Straight | | Circular | | |
|---|---|---|---|---|---|
| Valance | $E_{N_{max}}$ | $N_{max}$ | $E_{N_{max}}$ | $N_{max}$ | $|\Delta E_{Coul}|$ |
| 2 | -21031.76 | 164 | -23873.44 | 162 | 93.66 |
| 3 | -21621.44 | 112 | -24453.47 | 110 | 93.34 |
| 4 | -22136.09 | 85 | -24957.72 | 84 | 93.00 |

(Note: $\Delta E_{Coul} = [E_{N_{max}}(circular) - E_{N_{max}}(straight)]/l_B$)

## 5. Electrostatic Ring Closure Energy

Kunze and Netz [30] (and later employed by others [31, 32]) analytically formulated the energy difference between a straight line charge and its circular form (surrounded by counterions) on linearized Debye-Huckel level. Both the line and its ring shape energies are comprised of two factors, one is pure Coulomb energy and the other is exponential factor due to the screening. It is not possible to calculate the exponential factor for screening directly from this study as the study has been performed considering non-screened Coulomb interactions. If screening was considered then the Debye-Huckel potential would act between the charges as [33]

$$\Psi_{DH}(r_{ij}) = \frac{e^2}{4\pi\varepsilon_o\varepsilon_r k_B T} \frac{\exp[-\kappa r_{ij}]}{r_{ij}} = \left(\frac{E_{Coul}}{k_B T}\right)\exp[-k r_{ij}] \quad [11]$$

where $\kappa^{-1} = \dfrac{1}{\sqrt{4\pi l_B \rho_o}}$ is the Debye screening length (salt free), $\rho_o$ is the bulk concentration of monovalent counterion species and $r_{ij}$ is the distance between either a counterion and a macroion point charge or a counterion and another counterion. But in case of strong Coulomb coupling this mean field approach is not applicable as the electrostatic interaction is much stronger than thermal energy.

The difference of the pure Coulomb energy (per unit $k_B T$) is given by [30]

$$|\Delta E_{Coul}| = |E_{ring} - E_{rod}| = \tau^2 l_B L (1 - \ln\dfrac{\pi}{2}) \qquad [12]$$

Where $\tau$ is the line charge density and $L$ is the length of the line charge so that $2\pi R = L$, where R is the radius of the ring. The right hand side of equation 12 is independent of valence of counterions. For $L = 510$ Å and $\tau \approx 0.59$ e/Å one can calculate $|\Delta E| \approx 96.78 l_B$.

It has been seen before [24] that the Coulomb part of the 'ring closer energy' $\Delta E_{Coul}$ remained almost unchanged with the change of counterion valance for neutral cases (without overcharging). But in case of overcharging when maximal acceptances are considered for both straight and its circular shapes, it varies a little with overcharging counterion valance. The last column of table 2 shows the values of $\Delta E_{Coul}$ in turms of $l_B$ (= 30. 34 Å).

## 6. Concluding remarks

The goal of this paper is to shed light on a so far un-noticed phenomenon of cationic DNA solution in strong Coulomb coupling regime which can have subtle impact on the physico-chemical picture of interactions among DNAs and its solution. It is an experimental fact that highly charged biomolecules become naturally overcharged in cationic solutions. In this study it has been found that in strong Coulomb coupling environment when overcharged DNAs assume circular morphology some of the cations can get rid of their surfaces (leaving the DNA fully overcharged). This occurs because of energy minimization. The bent morphology of DNAs is energetically more favorable than its straight form. The freed cations then return to the solution and increase the cationic concentration of the solution. The number of freed cations varies with valance. The divalent and trivalent cations are found to be more prone to get rid off the overcharged DNA than tetravalent. Cationic solution concentration, therefore, can be appreciably high if the DNA concentration is significant and if the cations are divalent or trivalent. The increase in cationic concentration can enhance or even introduce a number of events in the solution. For example, the precipitation or reentrance of DNA [34, 35, 36] in cationic solution depends on specific concentration of added cations. Since static dielectric constant depends on ionic strength [36, 37, 38, 39] and electrostatic interaction is inversely proportional to dielectric constant, the increase in cationic concentration due to bending of DNA can reduce electrostatic attraction among precipitated DNAs to reenter into the solution. Increase of cationic concentration due to bending is an intermediate picture which can be considered in analytical models to explain some of the ionic concentration dependent features like the above one and similar others.

It has also been observed that the maximal acceptance varies with the amount of bending. For tetravalent cations only one ion gets rid of the DNA surface when the DNA bends to at least half cycle. While for divalent the same happens when the DNA is almost a circle and finally two ions leave the surface when the DNA is a complete cycle. But in case of trivalent one counterion gets freed at the very beginning of bending and finally leaves two when the DNA bends to a complete cycle. Thus, this study indicates a new phenomenon of possibility of enhancement of cationic concentration of a DNA solution that depends on the amount of DNA bending and valance of cations.

A simple and straightforward theoretical model (namely modified Sctchard approach) yields results which are in excellent agreement with all the simulation outputs. It confirms the accuracy of the calculations.

**Acknowledgement**